\documentclass[twocolumn]{revtex4}
\usepackage{graphicx}
\usepackage{dcolumn}
\usepackage{amsmath}

\newcommand{\beq}{\begin{equation}}
\newcommand{\eeq}{\end{equation}}

\begin{document}

\title{Ring dark solitons and vortex necklaces in Bose-Einstein condensates}
\author{G. Theocharis$^1$, D.J. Frantzeskakis$^{1}$, P.G. Kevrekidis$^2$, B.A.
Malomed$^3$ and Yuri S. Kivshar$^4$}

\begin{abstract}
We introduce the concept of ring dark solitons in Bose-Einstein condensates.
We show that relatively shallow rings are not subject to the snake
instability, but a deeper ring splits into a robust ring-like cluster of
vortex pairs, which performs oscillations in the radial and azimuthal
directions, following the dynamics of the original ring soliton.
\end{abstract}

\affiliation{ $^{1}$
Department of Physics, University of Athens,
Panepistimiopolis, Zografos, Athens 15784, Greece \\
$^{2}$ Department of Mathematics and Statistics,
University of Massachusetts, Amherst MA 01003-4515, USA \\
$^{3}$ Department of Interdisciplinary Studies, Faculty of
Engineering, Tel Aviv University, Tel Aviv 69978, Israel \\
$^4$ Nonlinear Physics Group, Research School of Physical Sciences
and Engineering, Australian National University, Canberra ACT
0200, Australia}

\maketitle


Intensive studies of Bose-Einstein condensates (BECs)
\cite{review} have drawn much attention to the dynamics of
nonlinear excitations such as bright \cite{bright} and dark
\cite{dark,dexp2,nsbec} solitons. In particular, dark solitons in
BECs were studied in detail \cite{motion}, and it was found that
they are subject to dynamical and thermal instabilities
\cite{dis}. The experimentally observed dynamical instability
\cite{dexp2} is due to their quasi-1D character: when embedded in
a higher dimension, a dark-soliton stripe becomes unstable against
transverse snaking \cite {nsbec,thsnbec}.

Transverse perturbations of dark solitons were extensively studied
in nonlinear optics (where such solutions exist in the case of
self-defocusing nonlinearity)\cite{si}. Since the wavenumber
band of the snake instability is limited by a maximum wavenumber
$Q_{\mathrm{m}}$, the instability can be suppressed by bending a
soliton stripe to close it into an annulus of length $L<2\pi
/Q_{m}$. The resulting \textit{ring dark solitons} (RDS's, i.e.,
annular troughs on a uniform background), first introduced in Ref.
\cite{KYR}, were studied in optics theoretically \cite{thring} and
experimentally \cite {expring1}. We note in passing that bright
solitons are unstable to collapse in higher dimensions \cite{si},
which also pertains to bright ring-shaped structures.

In this Letter we introduce the concept of RDS in BEC, as a novel class of
solitons which can be experimentally created by means of known
phase-engineering techniques \cite{expring1,expring2}. A principal
difference from optics is that the RDS dynamics in BEC is \emph{temporal},
while in optical media it is of the spatial type \cite{KYR}. Physical means
available to control BEC, such as dc and ac magnetic fields, are also
completely different from those employed in optics. Using the perturbation
theory \cite{kivpr} and simulations, we demonstrate that shallow RDS's in
BEC are long-lived objects, that may be observed experimentally on a
relevant time scale. On the contrary, deep RDS's are subject to snake
instability, splitting into ring-shaped vortex arrays (``vortex necklaces'')
which, eventually, reduce to four vortex-antivortex pairs, which perform
robust double-oscillatory motion in radial and azimuthal directions. All
these dynamical features are drastically different from those known in
optics, showing that the concept of RDS comprises a much broader range of
behaviors than it was known previously. Moreover, we establish a link of RDS
in BEC to ring fluxons in large-area Josephson junctions
\cite{Geicke,Maslov}.

The evolution of the BEC is governed by the Gross-Pitaevskii equation with a
trapping potential $V(\mathbf{r)}$ \cite{review}. We consider a disk-shaped
trap of the form $V(r,z)=m(\omega _{r}^{2}r^{2}+\omega _{z}^{2}z^{2})/2$,
where $r^{2}=x^{2}+y^{2}$, $m$ is the atom mass, $\omega _{r,z}$ are the
confinement frequencies in the radial and axial directions, and $\Omega
\equiv \omega _{r}/\omega _{z}\ll 1$. Then, following Refs. \cite{GPE1d},
one can derive an equation for a normalized mean-field wave function
$u(t,r)$:
\begin{equation}
iu_{t}=-(1/2)\left( u_{rr}+r^{-1}u_{r}\right) +|u|^{2}u+(1/2)\Omega
^{2}r^{2}u.  \label{GP}
\end{equation}

We now seek for solutions to Eq. (\ref{GP}) describing rings of lower
density on a background, which is described by the Thomas-Fermi (TF)
approximation. For Eq. (\ref{GP}), the latter is $u_{0}=\sqrt{\mu -\left(
1/2\right) \Omega ^{2}r^{2}}\exp \left( -i\mu t\right) $, where $\mu $ is
the chemical potential. As $\Omega $ is small, we can define a region where
the trapping potential is much smaller than $\mu $, then $u_{0}(r,t)\approx
\left[ \sqrt{\mu }-(1/4\sqrt{\mu })(\Omega r)^{2}\right] \exp (-i\mu t)$. To
describe the dynamics of RDS on top of the background $u_{0}$, we look for a
solution of Eq. (\ref{GP}) in the form $u\equiv u_{0}(r,t)\upsilon (r,t)$,
where the complex field $\upsilon (r,t)$ will introduce the ring soliton.
For $\Omega \ll 1$, the most interesting case is when the radius of the ring
is large enough, so that $1/r=O(\Omega )$. In this case, upon redefining
$t\rightarrow \mu t$, $r\rightarrow \sqrt{\mu }r$, Eq. (\ref{GP}) leads to an
effective perturbed nonlinear Schr\"{o}dinger (NLS) equation,
\begin{equation}
i\upsilon _{t}+(1/2)\upsilon _{rr}-(|\upsilon |^{2}-1)\upsilon =P(\upsilon ),
\label{pnls}
\end{equation}
where $P$ stands for the effective perturbation,
\begin{equation*}
P(\upsilon )\equiv \mu ^{-1}\left[ \left( 1-|\upsilon |^{2}\right) \upsilon
W(r)+(1/2)W^{\prime }(r)\upsilon _{r}-\sqrt{\mu }(2r)^{-1}\upsilon _{r}
\right] ,
\end{equation*}
with $W(r)\equiv (\Omega r)^{2}/2$, all terms in the perturbation $P$ being
on the same order of smallness.

We apply the perturbation theory for dark solitons \cite{kivpr} to Eq. (\ref
{pnls}). We start with the unperturbed dark soliton and seek for a ring-like
solution to Eq. (\ref{pnls}) as $\upsilon (r,t)=\cos \varphi (t)\cdot \tanh
\xi +i\sin \varphi (t)$, where $\xi \equiv \cos \varphi (t)\left[ r-R(t)
\right] $, and $\varphi (t)$ and $R(t)$ are slowly varying phase $(|\varphi
|<\pi /2)$ and radius of the ring soliton. It is straightforward to derive
perturbation-induced evolution equations:
\begin{equation}
\frac{d\varphi }{dt}=-\frac{\cos \varphi }{2\mu }\frac{dW}{dR}+\frac{\cos
\varphi }{3\sqrt{\mu }R},\;\;\;\;\;\;\frac{dR}{dt}=\sin \varphi \,.
\label{evo}
\end{equation}
Combining these, we arrive at an equation of motion for the RDS radius:
\begin{equation}
\frac{d^{2}R}{dt^{2}}=\left[ -\frac{1}{2}\frac{dW(R)}{dR}+\frac{1}{3R}\right]
\left[ 1-\left( \frac{dR}{dt}\right) ^{2}\right] ,  \label{eqm}
\end{equation}
in which we set $\mu \equiv 1$, as $\mu $ can be eliminated from Eqs. (\ref
{evo}) by the transformation $t\rightarrow \sqrt{\mu }t,\,\Omega \rightarrow
\sqrt{\mu }\Omega $.

In the limiting case of a plane soliton, $R\rightarrow \infty $, and with
$\cos \varphi \approx 1$, it is readily observed that Eq. (\ref{eqm}) reduces
to an equation of motion for the soliton's radius, $d^{2}R/dt^{2}+(\Omega
/2)R=0$, which recovers a known result for a quasi-1D dark soliton in a
parabolic potential \cite{motion}: it oscillates in an harmonic trap with
the frequency $\Omega /\sqrt{2}$. On the other hand, in the absence of the
trapping potential $V$ and for an almost black (deepest) soliton, Eq. (\ref
{eqm}) demonstrates that the curvature-induced effective potential is
$U=-(1/3)\ln R$, which recovers a result known in the context of nonlinear
optics \cite{KYR}. In the present case, a combination of the trapping
potential and ring curvature gives rise to an effective potential well for
the soliton's radial degree of freedom, $\Pi (R)=(1/2)(\Omega
R)^{2}-(1/3)\ln R$, which resembles oscillations of a circular sine-Gordon
(sG) kink in an axially symmetric anti-trap potential, which is possible in
large-area Josephson junctions \cite{Maslov} (without the anti-trap
potential, the circular sG kink periodically collapses and bounces back, forming
an extremely robust pulsating object \cite{Geicke})

The above consideration shows that RDS's can be found in BEC both as
oscillating rings and stationary ones, trapped at the bottom of the
potential well $\Pi (R)$, i.e., with the radius $R_{0}=\Omega ^{-1}\sqrt{2/3}
$. For the oscillatory states, the points $R_{\mathrm{min}}$ and $R_{\mathrm{
\ max}}$ between which $R(t)$ oscillates can be found, using Eqs. (\ref{evo}
) to eliminate $\sin ^{2}\varphi $:
\begin{equation*}
R_{\mathrm{min}}=\left[ -(2/3)w(0,\eta )\right] ^{1/2}\Omega ^{-1},R_{
\mathrm{max}}=\left[ -(2/3)w(-1,\eta )\right] ^{1/2}\Omega ^{-1},
\end{equation*}
where $w(k,\eta )$ is the Lambert's $w$-function defined as the inverse of
$\eta (w)=w\exp (w)$ \cite{lamb}, the integer $k$ is the branch number of the
function ($k=0$ corresponds to the principal value), and $\eta \equiv
-3W[R(0)]\cos ^{6}\varphi (0)\exp \left\{ -3W[R(0)]\right\} $.

The possible existence of the stationary and oscillating ring solitons is
specific to BECs, where they are supported by the trapping potential, while
their counterparts in nonlinear optics expand indefinitely \cite{KYR}.
Stability of the ring solitons, trapped at or around $R=R_{0}$, against
transverse perturbations should be tested in direct simulations.

Using the split-step Fourier method \cite{ta}, we integrated Eq. (\ref{GP})
numerically, with an initial configuration (IC) of the form
\begin{equation*}
u(r,0)=\left( 1-\Omega ^{2}r^{2}/4\right) \left[ \cos \varphi (0)\tanh
Z(r)+\sin \varphi (0)\right] ,
\end{equation*}
where $Z(r)=(r-R_{0})\cos \varphi (0)$, $\Omega =0.028$, $R_{0}=28.9$, and
$\cos \varphi (0)$ is the depth of the input soliton. The cases of
oscillating and stationary RDS can be considered, taking $\cos \varphi
(0)\neq 1$ and $\cos \varphi (0)=1$, respectively. Simulations verify that
both oscillating and stationary RDS's exist, and their dynamics can be
effectively described by Eq. (\ref{eqm}), up to a certain time. Then,
instabilities develop: RDS either slowly decays into radiation [for $\cos
\varphi (0)<0.67$], or, for $\cos \varphi (0)\geq 0.67$, snaking sets in,
leading to formation of vortex-antivortex pairs arranged in a robust
ring-shaped array (vortex cluster).

To illustrate these generic scenarios, we first take the case with $\cos
\varphi (0)=0.6$ and $\sin \varphi (0)=-0.8$. The corresponding initial
structure is shown in Figs. \ref{fig1}(a,b). According to the analytical
results, in this case RDS is expected to oscillate with the period $T=240$
between widely different limits, $R_{\mathrm{min}}=3.8$ and
$R_{\mathrm{max}}=69.7$, the latter being almost at the
rim of the BEC cloud, whose TF
radius is$\,\approx 70$. It is indeed observed that RDS initially shrinks,
attains the maximum contrast at $R=R_{\mathrm{min}}$ [Fig. \ref{fig1} (c)],
and bounces back. After reaching $R_{\mathrm{max}}$ and bouncing from it,
RDS starts to emit radiation in the form of shallow concentric dark rings,
as shown in Fig. \ref{fig1}(d). Due to the radiation loss, RDS becomes
shallower and, as a result, it accelerates, decreasing the period of the
oscillations. We observe that RDS performs at least three complete cycles of
the oscillations before final decay, which occurs at $t\approx 400$.
Qualitatively, this dynamical instability resembles that of a stripe
(rectilinear) dark soliton in BECs \cite{dis}.

\begin{figure}[tbp]
\centerline{
} \vskip
-0.1in
\caption{Evolution of a ring dark soliton (RDS) with $R(0)=28.9$ and
$\cos\protect\varphi(0)=0.6$. (a,b) The initial profile shown by cross-section
and gray-scale density plots, where RDS corresponds to a grey ring. (c) RDS
shrinks to the minimum radius (at $t=40$). (d) Beginning of the emission
of dark concentric rings (at $t=160$).}
\label{fig1}
\end{figure}

To translate the results into units relevant to the experiment \cite
{dark,dexp2}, we assume $^{87}$Rb condensate of radius $30\mu $m, containing
$20,000$ atoms in a disk-shaped trap with $\omega _{r}=2\pi \times 18$ Hz
and $\omega _{z}=2\pi \times 628$ Hz. In this case, the RDS considered above
has the radius $R_{0}=12.4\mu $m, it starts to emit radiation at $t\simeq 40$
ms, and finally decays at $t=100$ ms. This time scale is much larger than
the lifetime of the dark stripe observed in Refs. \cite{dark,dexp2}, hence
moderately shallow RDS's can be observed too.

Deep RDS's develop the snake instability, which results in the formation of
vortex pairs in multiples of four, namely $4$ [for $0.67\leq \cos \varphi
(0)<0.8$], $8$ [for $0.8\leq \cos \varphi (0)<0.9$], $12$ [for $0.9\leq \cos
\varphi (0)<0.95$], or $16$ [for $0.95\leq \cos \varphi (0)\leq 1$].
Originally, all the pairs are set along a single ring, creating a
necklace-like structure. The subsequent evolution of the necklaces is
characterized by a transient stage, when quartets of pairs are successively
expelled off the necklace, drift inward to the center of the condensate and
disappear there. Eventually, there remains a pattern consisting of precisely
four vortex pairs. They are arranged along a ring that slowly oscillates
between $R_{\mathrm{min}}$ and $R_{\mathrm{max}}$, i.e., the same limits
between which the initial RDS oscillated prior to the onset of the
instability. Simultaneously, the vortices and antivortices perform an
oscillatory motion along the ring, so that the configuration periodically
switches between x- and $+$-like shapes.

The robust necklace patterns consisting of vortex pairs resemble stable
clusters of globally linked vortices (of one sign, rather than of the
vortex-antivortex type) that were recently found in a 2D BEC model \cite
{Barcelona}; however, the number of vortices in those clusters could be
arbitrary (at least, $2$, $4$, and $8$). Another similar object are necklace
soliton clusters in nonlinear optics, which, however, are not stationary,
gradually expanding \cite{Moti} or rotating \cite{Des}.

\begin{figure}[tbp]
\caption{Evolution of RDS with $R(0)=28.9$ and $\cos \protect\varphi
(0)=0.76 $. (a) Spontaneous undulations developed on the ring at $t=60$. (b)
Four vortex pairs are formed, at $t=70$, as a result of the snake
instability.}
\label{fig2}\centerline{
} \vskip
-0.1in
\end{figure}

The double-oscillatory state persists for long times, typically up to
$t\approx 2000$ \textbf{(}which is $\simeq $ $500$ ms for the typical case
specified above); still later, due to significant distortion of the
condensate as a whole, the radial symmetry of the system breaks up,
resulting in eventual annihilation of all the vortex pairs. To illustrate
these scenarios, we display two cases, which correspond to situations where
the instability initially creates the minimum ($4$) or maximum ($16$) number
of vortex pairs.

First, we consider RDS with $\cos \varphi (0)=0.76$ and
$R_{\mathrm{min}}=8$, $R_{\mathrm{max}}=58$.
This IC is very similar to that in Fig.
\ref{fig1}(b). It initially shrinks and attains
$R=R_{\mathrm{min}}$ [as in Fig.
1(c)]. After bouncing and subsequently expanding to the rim of the BEC
cloud, it starts snaking, see Fig. \ref{fig2}(a), which is a precursor of
splitting. Finally, it splits into four vortex pairs, see Fig. 2(b). The
persistent quartet of the pairs arranges itself in a ring configuration. The
ring performs slow radial oscillations between $R_{\mathrm{min}}$ and $R_{
\mathrm{max}}$ with a period $T\approx 400$, which corresponds to $\simeq 100
$ ms, up to $t\approx 2000$. Simultaneously, the vortices and antivortices
move along the ring, so that they form an x-like configuration at $R=R_{
\mathrm{max}}$ [Fig. \ref{fig3} (a)], then an octagon [Fig. \ref{fig3}(b)],
and then a $+$-like pattern. As is seen in Fig. \ref{fig3}(c), the latter
one shrinks to $R=R_{\mathrm{min}}$, then it bounces and expands, attaining
$R=R_{\mathrm{max}}$ [Fig. \ref{fig3}(d)], and evolves into the x-like
pattern, and then the cycle repeats itself.

\begin{figure}[tbp]
\centerline{
} \vskip
-0.1in
\caption{Evolution of four vortex pairs created by the instability of RDS
from Fig. \ref{fig2} for: (a) $t=120$, (b) $t=240$, (c) $t=420$, and (d)
$t=540$. The vortex pairs remain on the ring and move along it, so that the
configuration oscillates between the x- and $+$-like configurations, while
the ring itself periodically shrinks and expands between $R_{\mathrm{min}}$
and $R_{\mathrm{\ max}}$.}
\label{fig3}
\end{figure}

Finally, we consider the evolution of a black ring soliton, with $\cos
\varphi (0)=1$, which, according to the analytical prediction, is expected
to be stationary [the corresponding initial state looks similar to that
shown in Fig. \ref{fig1}(b)]. First, this configuration indeed remains
stationary. However, Fig. \ref{fig4}(a) shows that, at $t=40$ ($\simeq $ $10$
ms), the ring starts to snake, which ends up with formation of a necklace
array of sixteen vortex pairs along the ring $R=R_{0}$, see Fig.
\ref{fig4}(b). The subsequent evolution of the necklace
results in annihilation of
eight pairs, which occurs in two steps. At first, four pairs drift inward,
where they disappear [Fig. \ref{fig4}(c)], leaving a nearly rectangular
array of twelve vortex pairs. Next, the $12$-pair pattern expunges two
quartets of vortex pairs. One quartet again moves inward and disappears near
the center, the other one drifts outward, while four vortex pairs stay at
$R=R_{0}$, see Fig. \ref{fig4}(c). Then, the four outward-moving vortex pairs
bounce from the rim of the condensate and move back inward, past the quartet
that stays put at $R=R_{0}$, and eventually disappear at the center. Thus,
there remains a pattern consisting of four vortex pairs, which still reside
at $R\approx R_{0}$, see Fig. 4(d) (due to the overall distortion of the
condensate at this late stage of the evolution, $R_{0}$ was properly
adjusted). The vortices and antivortices from these pairs move along the
ring, so that the configuration performs very slow oscillations between the
x- and $+$-like shapes. We have observed almost three complete cycles of
such oscillations with the period $T\approx 500$ ($\simeq $ $125$ ms).

\begin{figure}[tbp]
\centerline{
} \vskip
-0.1in
\caption{Evolution of an original stationary ring soliton with $\cos\protect
\varphi (0)=1$ and $R(0)=R_{0}=28.9$. The initial configuration, which is
very similar to that in Fig. \ref{fig1}(b), remains undistorted up to
$t=40$. (a) Undulations of RDS at $t=40$. (b) Sixteen vortex pairs, formed by
$t=60 $ as a result of the snake instability. Further snapshots are shown for
(c) $t=240$ and (d) $t=760$. The initial set of $16$ vortex pairs reduces
first to $12$, then to $8$, and eventually to $4$ pairs, through successive
annihilations of two quartets at the center of the condensate. Four
surviving vortex pairs stay at $r=R_{0}$, oscillating between the x- and
$+$-like patterns.}
\label{fig4}
\end{figure}

In conclusion, we have introduced the concept of ring dark solitons in
Bose-Einstein condensates, and predicted the existence of both oscillatory
and stationary solitons. Simulations show that perturbation theory
accurately describes the unperturbed RDS dynamics. However, instabilities
gradually set in and, as a result, shallow RDS's slowly decay, while deeper
ones develop the snake instability. In the latter case, a necklace array
consisting of vortex-antivortex pairs appears, the number of pairs being a
multiple of $4$. Eventually, it relaxes to a set of four pairs which sit on
a ring oscillating in the radial direction between the same limits which
confined the oscillations of the original RDS; simultaneously, the pairs
perform oscillatory motion along the ring.

This work was supported by the Special Research Account of the University of
Athens (GT, DJF), a UMass FRG and NSF-DMS-0204585 (PGK), Binational
(US-Israel) Science Foundation under grant No. 1999459 (BAM), and the
Australian Research Council (YSK). Discussions with H.E. Nistazakis are
gratefully acknowledged.

\end{document}